\documentclass[%
 reprint,
 amsmath,amssymb,
 aps,
]{revtex4-2}
\usepackage{graphicx}
\usepackage{dcolumn}
\usepackage{bm}
\usepackage{xcolor}
\usepackage{hyperref}
\usepackage[normalem]{ulem}
\usepackage{mhchem}

\usepackage{braket}
\begin{document}

\preprint{APS/123-QED}

\title{
Radical Pair Mechanism and the Role of Chirality-Induced Spin Selectivity during Planaria Regeneration: Effect of Weak Magnetic Field on ROS levels}  
\author{Yash Tiwari}
\affiliation{%
Department of Electronics and Communication, Indian
Institute of Technology, Roorkee, India 
}%
\author{Parul Raghuvanshi}
\affiliation{%
Department of Electronics and Communication, Indian
Institute of Technology, Roorkee, India 
}%
\author{Vishvendra Singh Poonia}%
\email{vishvendra@ece.iitr.ac.in}
\affiliation{%
Department of Electronics and Communication, Indian
Institute of Technology, Roorkee, India 
}%

\date{\today}

\begin{abstract}
Planarian is an intriguing model system wherein the effect of electric and magnetic fields can be studied on various biochemical pathways during cell morphogenesis. Recent experimental observations have demonstrated the non-trivial modulation of reactive oxygen species (ROS) levels by weak magnetic field during planaria regeneration. However, the underlying biophysical mechanism behind this remains elusive. 
In this paper, we study the radical pair mechanism to explain the effect of weak magnetic fields on ROS modulation during planaria regeneration to explain the experimental results. 
We also investigate the effect of chirality-induced spin selectivity (CISS) on ROS levels by including it in the framework of the radical pair mechanism. 
We conclude that the inclusion of CISS not only explains the experimental results better but also allows for the radical pair model to have more parametric space to satisfy the experimental constraints. This study explains the crucial process of ROS modulation by the weak magnetic field with and without CISS thereby paving the way to unravel the vast domain of ROS modulation for desired outcomes.
\end{abstract}

\maketitle

\section{\label{INTRODUCTION}Introduction}
Several research studies emphasize the significant impact of electromagnetic fields on diverse biological processes and systems \cite{levin2003bioelectromagnetics,polk1995handbook,shupak2003therapeutic}. Morphogenesis, a biological process, is affected by electromagnetic fields, with planarians serving as a notable model for comprehending morphogenesis and cell communication. Planarians, recognized for their exceptional regenerative abilities, are often studied in this context ~\cite{levin2019planarian,lobo2012modeling,salo2009planarian,reddien2018cellular,gentile2011planarian,umesono2013molecular}.
Planaria's impressive regenerative ability is due to the presence of a set of robust adult stem cells known as neoblasts. Once injured, these cells migrate to the injured site and form a specialized structure known as a blastema. The blastema subsequently undergoes multiplication and transformation, regenerating the lost body parts like the head, trunk, or tail \cite{wenemoser2010planarian,scimone2014neoblast,alvarado2005multicellularity,rink2013stem,reddien2004fundamentals}. The formation of the blastema is influenced by specific signaling molecules known as reactive oxygen species (ROS). Blocking and activation of these molecules contribute to the change in the formation of blastema. Consequently, this impacts the growth of new tissues and affects the process of planarian regeneration \cite{pirotte2015reactive,jaenen2021reactive,kinsey2023weak}.
Recent insights into how living organisms interact with electromagnetic radiation indicate the possibility of discovering novel techniques for controlling ROS within the body using weak magnetic fields (WMFs)~\cite{kinsey2023weak}. However, the exact biophysical mechanism behind the effect of weak magnetic fields on ROS levels is not fully known. Radical pair mechanism has been proposed to elucidate the impact of weak magnetic fields on modulation of ROS levels~\cite{barnes2015effects,barnes2018role,barnes2020setting,rishabh2022radical,wang2006zeeman,zadeh2022magnetic,smith2021radical,nair2023radical}.\\
More generally, the radical pair mechanism (RPM) has emerged as a prominent theory explaining the effect of magnetic field on various biological and chemical systems~\cite{zadeh2022magnetic,jones2016magnetic,fan2021analgesic,lambert2013che,mohseni2014quantum} 
In particular, it has been studied in detail for the cryptochrome protein present in bird's retina and is considered a potential explanation for the avian magnetoreception \cite{hiscock2016quantum,fay2020quantum,hore2016radical,ball2011physics,lee2014alternative}.
Moreover, it has recently been reported that CISS might play a role in conjunction with radical pairs in birds for the navigational compass \cite{tiwari2022role,tiwari2023quantum,luo2021chiral}. The radical pair mechanism involves electron transport steps during the formation and recombination~\cite{hore2016radical,ritz2000model,smith2022driven}. This involves the movement of electrons through protein molecules. Owing to the chirality of protein molecules, the chiral-induced spin selectivity (CISS) effect could play an essential role in the electron transport part of the reaction. Although the exact role of CISS in the avian magnetoreception has been debated, there are strong evidence in support of its presence in various biochemical reactions involving electron transfer or rearrangement in chiral molecules~\cite{naaman2022chiral,naaman2020chiral,bloom2022homochirality,kumar2017chirality,mondal2016spin}. The origin of CISS is attributed to the spin-orbit interaction and the electrostatic potential provided by the chiral molecules~\cite{dalum2019theory,michaeli2019origin,matityahu2016spin,gohler2011spin,naaman2012chiral,naaman2015spintronics,xu2021magnetic,wong2021cryptochrome}.   \\
In this work, we examine the radical pair mechanism in light of recent experimental work pertaining to ROS level modulation by the weak magnetic field~\cite{van2019weak}. Furthermore, we also investigate the effect of chirality-induced spin selectivity (CISS) in the modulation of Reactive Oxygen Species (ROS) in the planarian system. To achieve this, we establish a theoretical model of the CISS-assisted radical pair mechanism, aligning our simulated results with experimental data \cite{kinsey2023weak}. Our investigation of the CISS-assisted radical pair mechanism is structured around three key considerations: 1) the absolute yield of the reaction, 2) the ratio of yield values at the experimental point of interest, and 3) the number of nuclei configurations that adhere to the experimental trend. Additionally, we explore two scenarios: 1) CISS presence exclusively during the formation of radicals and 2) CISS presence during both the formation and recombination of radical pairs.
The manuscript has been organized as follows: Section~\ref{METHODOLOGY} discusses the simulation methodology followed for analysis. Section ~\ref{RESULTS}  discusses the results, where subsection~\ref{RP_noCISS} discusses the RP model with no CISS  subsection~\ref{RP_CISS} discusses the RP model with CISS, subsection~\ref{RP_CISS_Rate} explores the impact of singlet and triplet recombination rate on planaria regeneration and ~\ref{RP_CISS_HighNuceli} discuss a system with higher number of nuclei. Finally, we discuss the shortcomings and conclusions of the study. 
\section{\label{METHODOLOGY}Methodology}

In the radical pair model of the planaria regeneration, an electron is excited in the acceptor molecule, creating a vacancy in the ground state. Another electron from a neighboring donor molecule travels in the chiral medium to fill this vacancy. It results in the formation of a radical pair. The spin 
operator of the electron on the donor molecule is $\hat{S}_{Dz}$ and on the acceptor molecule is $\hat{S}_{Az}$. Therefore, the spin state of the above-formed radical pair is governed by the following Hamiltonian~\cite{cintolesi2003anisotropic,fay2020quantum,luo2021chiral}
\begin{equation} 
\label{Hamiltonian_RPM}
\begin{split}
\hat{H}  = \omega.(\hat{S}_{Az}+\hat{S}_{Dz}) + \sum_{i\in{D,A}}\sum_{k}\hat{S}_{i}.A_{ik}.\hat{I}_{ik} \\ -
J(2\hat{S}_{A}.\hat{S}_{D} + 0.5) +
\hat{S}_{A}.D.\hat{S}_{B}
\end{split}
\end{equation}
where $\omega=g\Bar{\mu_B} \Bar{B}$, 
$\Bar{B} = B_0\Bar{z}$ where $B_0$ corresponds to the applied magnetic field. $J$ and $D$ are the exchange and dipolar interactions. $A$ is the hyperfine tensor depicting interactions between electrons and neighboring nuclear spins.
\begin{figure}
\centering
\includegraphics[width=90mm,keepaspectratio]{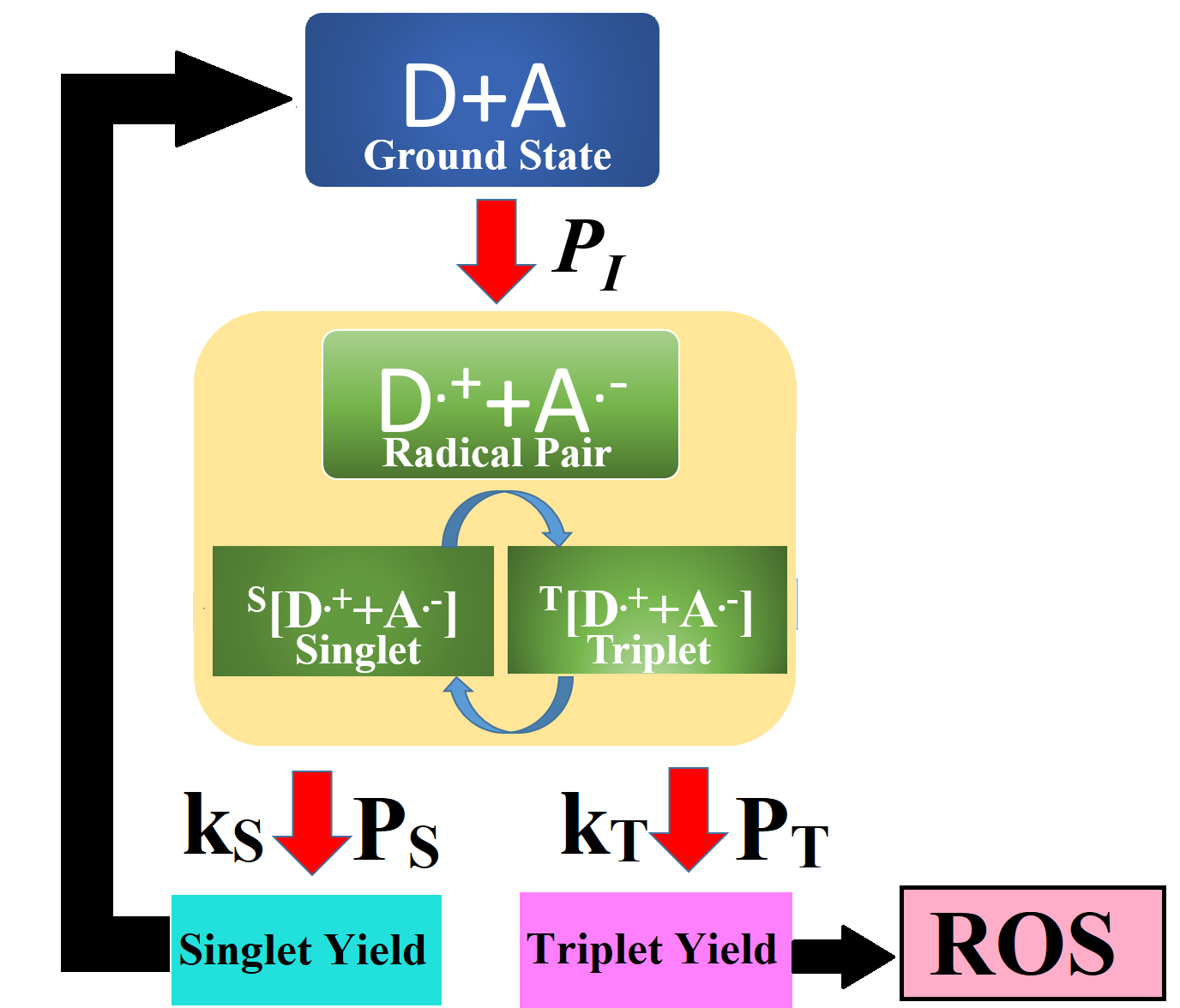}
\caption{The schematic for the CISS-assisted radical pair mechanism for planaria regeneration where $D$ denotes the donor molecule, $A$ represents the acceptor molecule. $D^{.+}$ is the donor radical, $A^{.-}$ is the acceptor radical. $k_S$ is the singlet recombination rate constant, and $k_T$ is the triplet recombination constant. The red arrows represent the role of CISS in the reaction pathways since it involves electron transport.}
\label{RECOMBINATION1}
\end{figure}

\begin{figure}
\centering
\includegraphics[width=90mm,keepaspectratio]{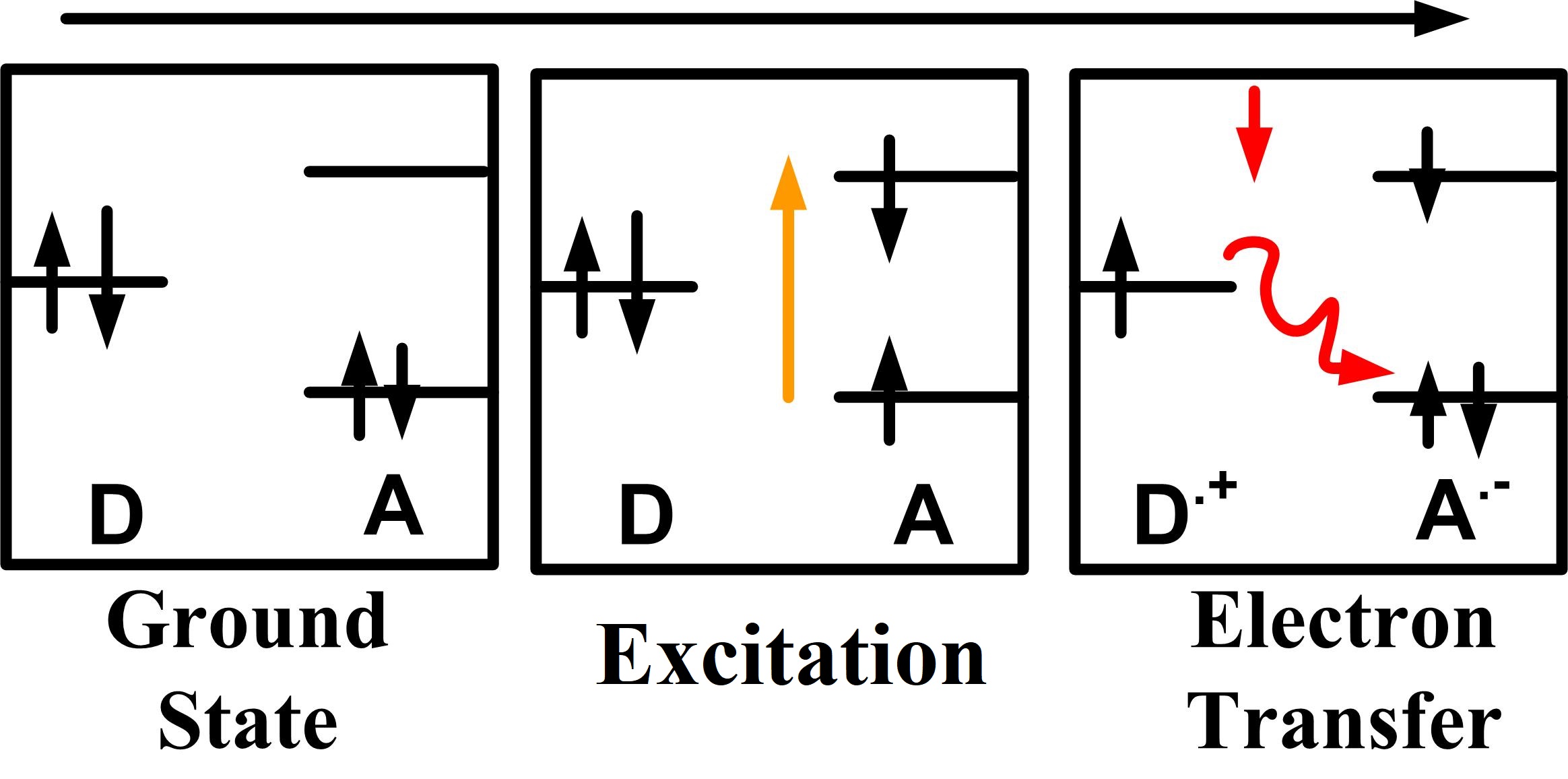}
\caption{
The radical pair formation begins with the donor molecule (D) and the acceptor molecule (A) in their ground states. The reaction initiates with the excitation of the acceptor molecule, causing an electron to transition to a higher energy state and leaving a vacancy in the ground state of the acceptor molecule. Only a specific electron spin movement occurs when the surrounding medium is fully chiral. This movement involves the transfer of an electron from the ground state of the donor molecule to fill the vacancy in the ground state of the acceptor molecule. This leads to the formation of the radical pair on donor ion $D^{.+}$ and acceptor ion $A^{.-}$.  }
\label{Formation_Ground_Pair}
\end{figure}

\begin{figure}
\centering
\includegraphics[width=90mm,keepaspectratio]{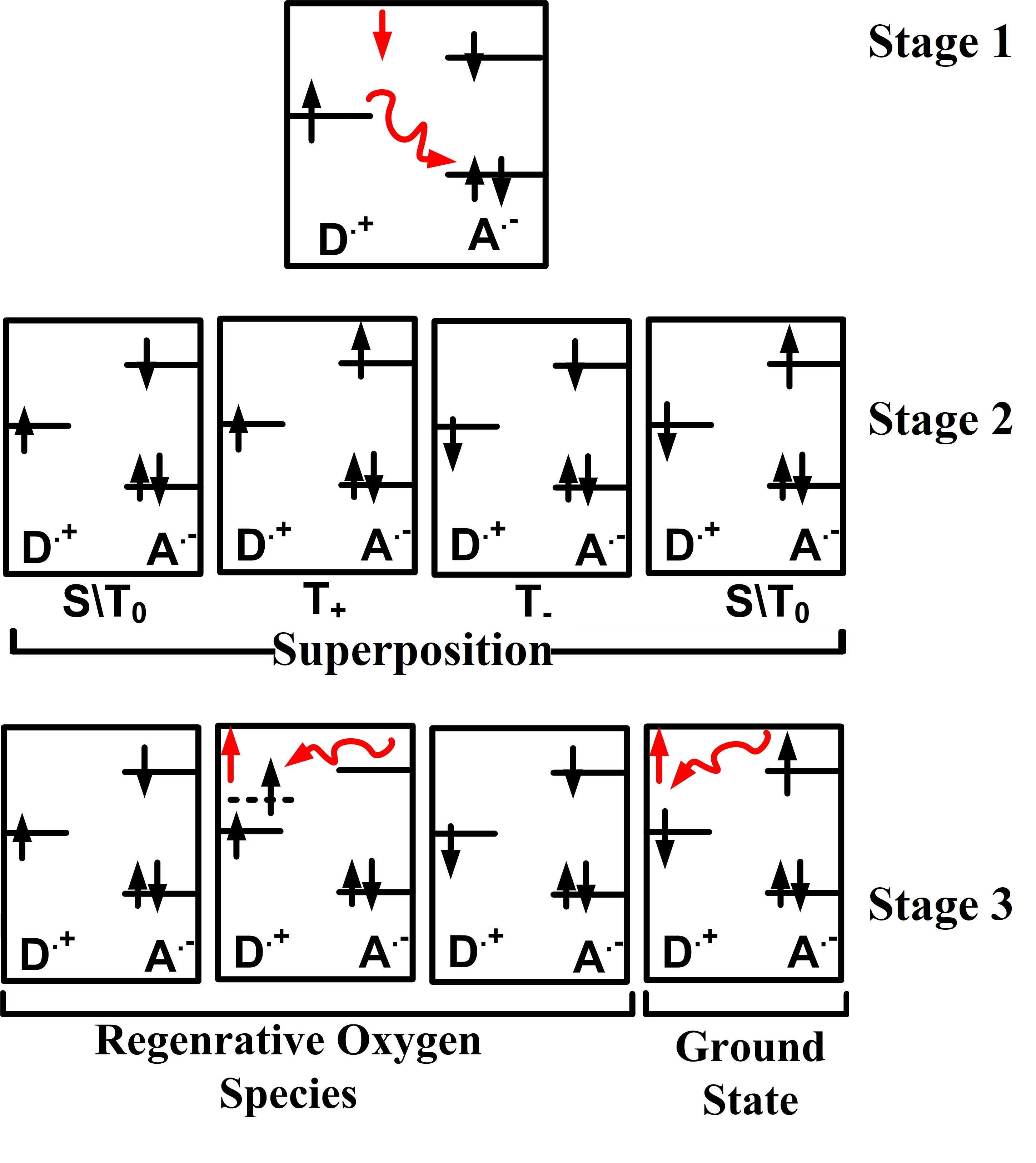}
\caption{The recombination of the radical pair involves the transition from the donor ion $D^{.+}$ and acceptor ion $A^{.-}$ to their respective ground states, as well as the generation of regenerative oxygen species. This process occurs in three stages. Stage 1 is the initial stage where the radical pair is formed. The chirality of the medium allows only the $\ket{\downarrow}$ state to move from the donor ion $D^{.+}$ to the acceptor ion $A^{.-}$. In Stage 2, the system exists in a superposition of singlet and triplet states. The spins of the isolated electrons on $D^{.+}$ and $A^{.-}$ change due to their interaction with the nuclei (hyperfine interaction). In stage 3, because the forward movement allowed the $\ket{\downarrow}$ state, the $\ket{\uparrow}$ state will move in the opposite (backward) direction for recombination.   }
\label{Recombination_Operator1}
\end{figure}

The spin state of the radical pair evolves under the zeeman and hyperfine interactions. Along with this evolution, the radical pair also recombines back to the singlet yield and triplet yield shown in Fig.~\ref{RECOMBINATION1}. The singlet yield corresponds to the ground state while the triplet state corresponds to reactive oxygen species(ROS) concentrationFig.~\ref{RECOMBINATION1}). 
The radical pair formation begins with the donor molecule (D) and the acceptor molecule (A) in their ground states. The reaction initiates with the excitation of
the acceptor molecule, causing an electron to transition to a higher energy state and leaving a vacancy in the ground state of the acceptor molecule. Electron transfer occurs from the ground state of the donor molecule to fill the vacancy in the ground state of the acceptor molecule. This leads to the formation of the radical pair on donor ion $D^{.+}$ and acceptor ion $A^{.-}$. Fig,\ref{Formation_Ground_Pair} illustrates the formation of the radical pair in a chiral medium where the spin state of the electron $\ket{\downarrow}$ is allowed in the forward propagation of the electron. In a non-chiral case, both $\ket{\downarrow}$  and $\ket{\uparrow}$ will be allowed in the formation of the radical pair.   \\
The recombination of the radical pair involves the transition from the donor ion $D^{.+}$ and acceptor ion $A^{.-}$ to their respective ground states, as well as the generation of regenerative oxygen species. This process occurs in three stages. Stage 1 is the initial stage where the radical pair is formed. The chirality of the medium allows only the $\ket{\downarrow}$ state to move from the donor ion $D^{.+}$ to the acceptor ion $A^{.-}$ as illustrated in Fig.\ref{Formation_Ground_Pair}. In Stage 2, the system exists in a superposition of singlet and triplet states. The spins of the isolated electrons on $D^{.+}$ and $A^{.-}$ change due to their interaction with the nuclei (hyperfine interaction). This stage is affected by the system Hamiltonian (Eq.\ref{Hamiltonian_RPM}). In stage 3, because the forward movement allowed the $\ket{\downarrow}$ state, the $\ket{\uparrow}$ state will move in the opposite (backward) direction for recombination. This recombination yield leads to the formation of the respective singlet and triplet yield. The triplet yield would contribute towards the ROS signaling, whereas the singlet yield would contribute towards the ground state of the donor and acceptor molecule.
In Fig.\ref{Recombination_Operator1}, we have illustrated it for a fully chiral medium and how chirality affects the formation of the yield. Due to chirality, since only $\ket{\uparrow}$ is allowed, we observe that yield formation of the yield of triplet state $\ket{\downarrow\downarrow}$ 
is inhibited.

The CISS effect plays a role in forming (Fig.\ref{Formation_Ground_Pair}) and recombining (Fig.\ref{Recombination_Operator1}) the radical pair as it involves electron transport through the chiral protein molecule. Therefore, the action of CISS is captured by the initial state $P_I$ and recombination state $P_S$ and $P_T$, shown with red arrows in Fig.~\ref{RECOMBINATION1}. 

\begin{equation} 
\label{INITIAL_STATE}
\begin{split}
\ket{\psi_{In}}=\frac{1}{\sqrt{2}}cos(\frac{\chi}{2})\ket{S}+sin(\frac{\chi}{2})\ket{T_0}
\end{split}
\end{equation}
\begin{equation} 
\begin{split}
\label{CISS_Intial_op}
P_{In}=\ket{\psi_{In}}\bra{\psi_{In}}=
\begin{bmatrix}
0 & 0 & 0 & 0\\
0 & 0.5(1-sin\chi) & -0.5cos\chi & 0\\
0 & -0.5cos\chi & 0.5(1-sin\chi) & 0\\
0 & 0 & 0 & 0
\end{bmatrix}
\end{split}
\end{equation}

Then the initial state density matrix is given as: $P_{Initial}= P_{In}\otimes \frac{I}{Z}$, where $\frac{I}{Z}$ corresponds to the mixed state of nuclei, and $Z$ is the size of the nuclear Hilbert space. The singlet recombination operator $P_S= {\ket{\psi_S}\bra{\psi_S}}$ accounts for recombination to the ground state where:

\begin{equation} 
\label{Singlet_Recombination_State}
\begin{split}
\ket{\psi_{S}}=\frac{1}{\sqrt{2}}cos(\frac{\chi}{2})\ket{S}-sin(\frac{\chi}{2})\ket{T_0}
\end{split}
\end{equation}
\begin{equation} 
\begin{split}
\label{Singlet_Recombination_Op}
P_{S}=\ket{\psi_{S}}\bra{\psi_{S}}=
\begin{bmatrix}
0 & 0 & 0 & 0\\
0 & 0.5(1+sin\chi) & -0.5cos\chi & 0\\
0 & -0.5cos\chi & 0.5(1-sin\chi) & 0\\
0 & 0 & 0 & 0
\end{bmatrix}
\end{split}
\end{equation}
The triplet recombination operator $P_T= P_{T+} + P_{T-} + P_{T0} $ accounts for recombination to the reactive oxygen species concentration. $P_{T+}$ correspond to triplet state when net magnetic moment $m_S = 1 $. $P_{T-}$ correspond to triplet state when net magnetic moment $m_S = -1 $ and $P_{T0}$ correspond to triplet state when net magnetic moment $m_S = 0 $

\begin{equation} 
\label{Triplet0_Recombination_State}
\begin{split}
\ket{\psi_{T0}}=\frac{1}{\sqrt{2}}sin(\frac{\chi}{2})\ket{S}+cos(\frac{\chi}{2})\ket{T_0}
\end{split}
\end{equation}
\begin{equation} 
\begin{split}
\label{Triplet0_Recombination_Op}
P_{T0}=\ket{\psi_{T0}}\bra{\psi_{T0}}=
\begin{bmatrix}
0 & 0 & 0 & 0\\
0 & 0.5(1-sin\chi) & 0.5cos\chi & 0\\
0 & 0.5cos\chi & 0.5(1+sin\chi) & 0\\
0 & 0 & 0 & 0
\end{bmatrix}
\end{split}
\end{equation}

\begin{equation} 
\begin{split}
\label{TripletPM_Recombination_Op}
P_{T+}=
\begin{bmatrix}
1 & 0 & 0 & 0\\
0 & 0 & 0 & 0\\
0 & 0 & 0 & 0\\
0 & 0 & 0 & 0
\end{bmatrix}
P_{T-}=
\begin{bmatrix}
0 & 0 & 0 & 0\\
0 & 0 & 0 & 0\\
0 & 0 & 0 & 0\\
0 & 0 & 0 & 1-sin\chi
\end{bmatrix}
\end{split}
\end{equation}

The CISS parameter $\chi\in [0,\frac{\pi}{2}]$ depends on the spin selectivity of the protein medium; $\chi = 0$ corresponding to no CISS and $\chi = \pi/2$ corresponding to the full CISS. The master equation governing the state evolution of the system is given as: 
\begin{equation} 
\label{MASTER_EQUATION}
\frac{d\hat{\rho}}{dt} =-i[\hat{H},\hat{\rho}(t)]-\frac{1}{2}k_S\{P_S,\hat{\rho}(t)\}-\frac{1}{2}k_T\{P_T,\hat{\rho}(t)\}
\end{equation}
 Where $k_S$ is the singlet recombination rate, and $k_T$ is the triplet recombination rate. $[A, B] = AB-BA$ correspond to the commutator whereas, $\{A, B\} = AB+BA$ is the anti-commutator.

\section{\label{RESULTS}Results}

As the literature states, the triplet yield is directly related to reactive oxygen species (ROS) concentration \cite{kinsey2023weak,van2019weak}. Consequently, it is crucial to determine how the distribution of triplet yield varies concerning the external magnetic field. Defining the yield product of the triplet state ($\phi_T$) is imperative; therefore, it is defined according to Eq.\ref{YIELD_TripletState}.

\begin{equation} 
\label{YIELD_TripletState}
\begin{split}
\phi_T=k_T\int_{0}^{\infty} Tr[\hat{\rho(t)}\hat{P_T}]dt
\end{split}
\end{equation}
$\hat{\rho(t)}$ is the solution of the master equation Eq.~\ref{MASTER_EQUATION}, $Tr$ is the trace over the state density matrix $\rho$. As reported in \cite{kinsey2023weak,van2019weak}, the ROS yield follows the following distribution

\begin{equation} 
\label{Experimental_YIeld_Condtions}
\begin{split}
ROS~at~B_z=45 \mu T > ROS~at~B_z=200 \mu T \\
ROS~at~B_z=200 \mu T< ROS~at~B_z=500 \mu T \\
ROS~at~B_z=45 \mu T < ROS~at~B_z=500 \mu T
\end{split}
\end{equation}

Hence, if RPM dictates the generation of reactive oxygen species (ROS), a plot of triplet yield ($\phi_T$) against the external magnetic field ($B_0$) should exhibit a strong correlation with the aforementioned experimental findings. In \cite{kinsey2023weak, van2019weak}, it is documented that the two nuclei involved may be an oxide ($O_2^{\dot -}$) and tryptophan ($Trp^{\dot +}$). The oxygen nucleus in $O_2^{\dot -}$ is nonmagnetic and, thus, does not contribute to the hyperfine interaction. Conversely, tryptophan contains multiple magnetic nuclei of nitrogen and hydrogen.
Considering the biological context where distances between nuclei and electrons are variable, a parametric approach has been adopted, specifically addressing the hyperfine interaction of tryptophan with an electron.

\subsection{Radical Pair model with no CISS for planaria regeneration}
\label{RP_noCISS}

In our initial investigation into the feasibility of the radical pair mechanism, we have defined three quantities that will assist us in our investigation. 
\begin{equation} 
\label{L_iparameters}
\begin{split}
    L_1=\frac{\phi_T(45 \mu T)}{\phi_T(200 \mu T)};
    L_2=\frac{\phi_T(500 \mu T)}{\phi_T(200 \mu T)};
    L_3=\frac{\phi_T(500 \mu T)}{\phi_T(45 \mu T)}
\end{split}
\end{equation}
Based on the experimental results reported in \cite{kinsey2023weak, van2019weak}, we expect parameters $L_1, L_2, L_3$ to have values at least greater than 1. We initially perform analysis considering there is no CISS ($\chi=0$) in the system and try to ascertain the planaria response to the external magnetic field.

\begin{figure}
\centering
\includegraphics[width=90mm,keepaspectratio]{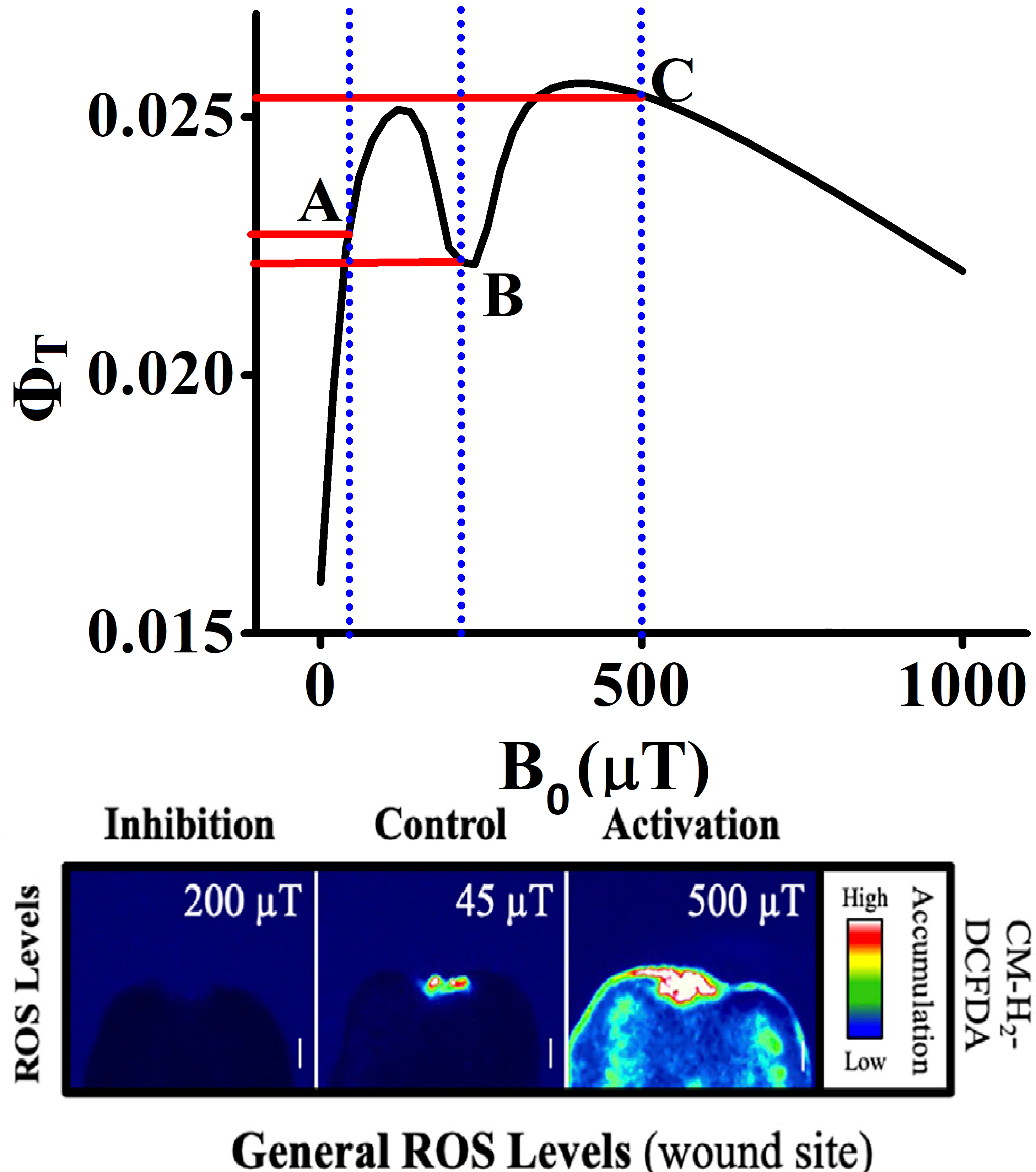}
\caption{Triplet Yield $\phi_T$ with respect to external magnetic field $B_0$  for  $\chi =0$ at  $k_S=10^8$ and $k_T=10^6$ with $D=0$ and $J=0$ at  hyperfine configuration $[a_{xx}, a_{yy}, a_{zz}] = [0.54, 0.06, 0.24 ] mT$. The three important points on the plot labeled as $A, B, C$ are our focus points since these points correlate to the ROS yield given in experimental data. Point $A$ correspond to $\phi_T(45\mu T)$, $B$ correspond to $\phi_T(200\mu T)$ and $C$ correspond to $\phi_T(500\mu T)$ }
\label{Figure1}
\end{figure}
In Figure \ref{Figure1}, the triplet yield ($\phi_T$) is plotted against the external magnetic field ($B_0$) for $\chi =0$ (no CISS) at $k_S=10^8$ and $k_T=10^6$, with $D=0$ and $J=0$, under the hyperfine configuration $[a_{xx}, a_{yy}, a_{zz}] = [0.54, 0.06, 0.24 ] mT$. The figure highlights points of interest corresponding to experimental conditions in planaria: $A$ for $\phi_T$ at $45\mu T$, $B$ for $\phi_T$ at $200\mu T$, and $C$ for $\phi_T$ at $500\mu T$. The experimental plot from references \cite{kinsey2023weak, van2019weak} is displayed at the bottom of Figure \ref{Figure1}, demonstrating the concentration of reactive oxygen species (ROC) in the experimental outcome.\\
A parametric analysis was conducted for all values of $[a_{xx}, a_{yy}, a_{zz}]$, with each component ranging from 0 to 3 $mT$ (step size 0.06 $mT$), ensuring that all three parameters ($L_1, L_2, L_3$) yield values greater than unity. Out of 132,651 possible hyperfine tensors, approximately 39 values met the desired condition. From Figure \ref{Figure1}, the observed values were $L_1=1.132, L_2=1.131, L_3=1.001$. Although ideally, a higher value of these ratios is preferred, the range of these ratios for all 39 possible hyperfine tensors was between 1-1.25. The maximum values of hyperfine tensor $[a_{xx}, a_{yy}, a_{zz}]$, where $L_1, L_2, L_3$ are greater than unity, were found to be $[0.54, 0.06, 0.24]mT$ . Consequently, the desired trend was only observed for low hyperfine values. Additionally, it was noted that the absolute value of $\phi_T$ was relatively low.\\
In the next section, chirality will be introduced to investigate whether spin selectivity provides a more accurate model that closely aligns with the experimental outcome.

\subsection{Radical Pair model with CISS for planaria regeneration}
\label{RP_CISS}
This section introduces spin selectivity arising from chirality, taking into account the nonzero value of $\chi$. Initially, we focus on spin selectivity during the formation of the radical pair from the acceptor and donor molecules. Consequently, the initial density matrix, denoted as $\hat{P}_{In}$, is influenced by variations in the CISS parameter $\chi$. However, the recombination operators are observed under the no CISS case ($\chi=0$), representing the standard singlet and triplet projection operators. In the subsequent part, we explore the role of CISS in both the formation and recombination of the radical. Thus, $\chi$ will impact not only the initial density matrix  $\hat{P}_{In}$ but also recombination operators $\hat{P}_{S}$ and $\hat{P}_{T}$. This work calculates the result for the fixed orientation of the CISS axis to the hyperfine z-axis \cite{luo2021chiral}.

\subsubsection{CISS only at the formation of radical pair }

\begin{figure}
\centering
\includegraphics[width=90mm,keepaspectratio]{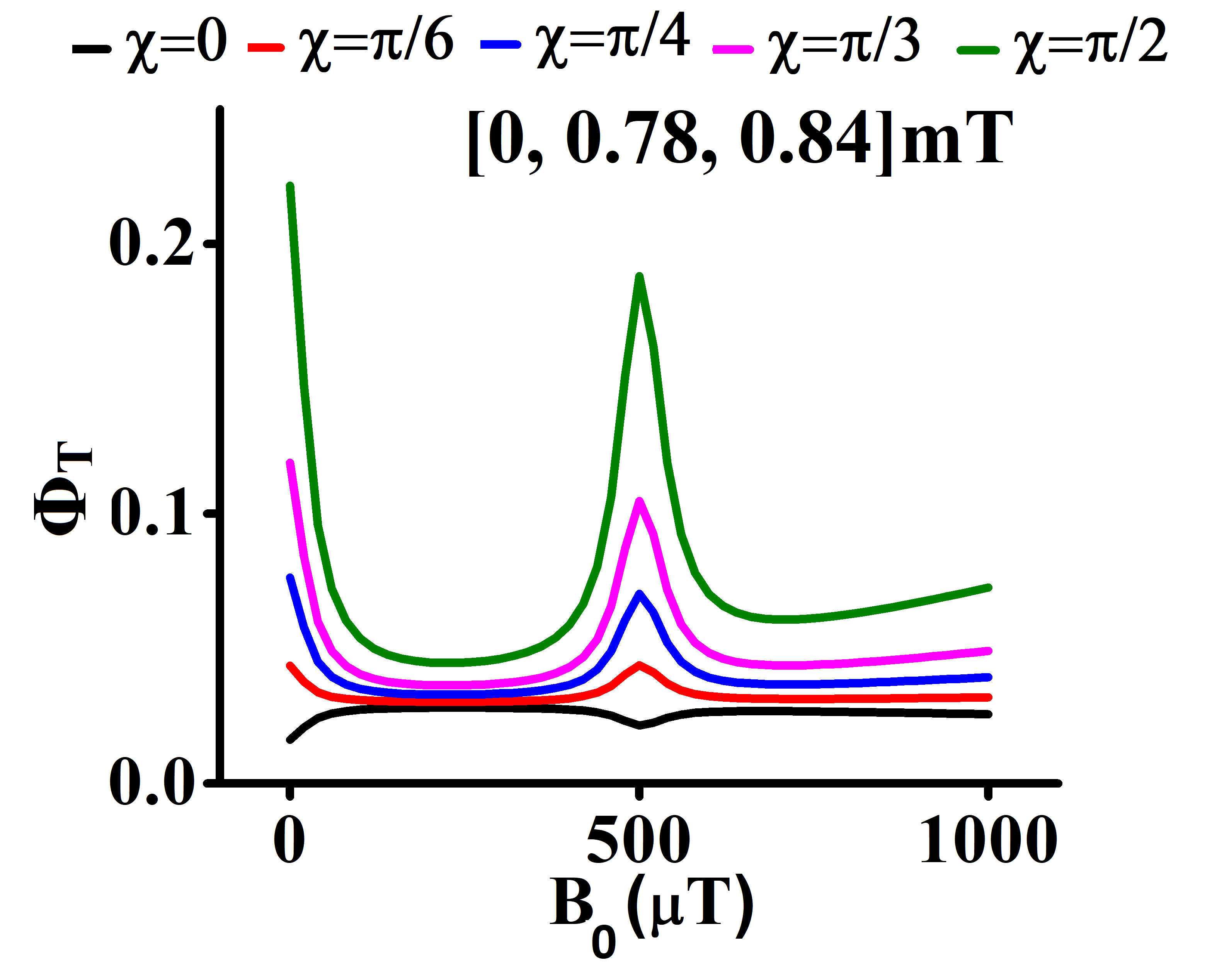}
\caption{Triplet Yield $\phi_T$ with respect to external magnetic field $B_0$  for five values of  $\chi =0, \frac{\pi}{6}, \frac{\pi}{4}, \frac{\pi}{3}, \frac{\pi}{2}$ at  $k_S=10^8$ and $k_T=10^6$ with $D=0$ and $J=0$  at hyperfine configuration $[a_{xx}, a_{yy}, a_{zz}] = [0, 0.78, 0.84 ] mT$ when CISS is present only in formation of the radical. }
\label{Figure3}
\end{figure}

In Figure \ref{Figure3}, the triplet yield ($\phi_T$) is graphed against the external magnetic field for five different values of $\chi =0, \frac{\pi}{6}, \frac{\pi}{4}, \frac{\pi}{3}, \frac{\pi}{2}$. This analysis is conducted at $k_S=10^8$ and $k_T=10^6$, with $D = 0$ and $J = 0$, and under the hyperfine configuration $[a_{xx}, a_{yy}, a_{zz}] = [0, 0.78, 0.84] mT$. Notably, CISS is present only during the formation of the radical in this scenario. We list the values of $L_1, L_2, L_3$ in Tab.\ref{table1} for Fig \ref{Figure3}.

\begin{table}
\centering
\caption{\label{table1}%
 $\phi_T$ for radical pair model based on ($O_2^{\do} -Trp^{\dot +}$ with one nuclei. The nuclei have spin one and hyperfine interaction as $[a_{xx}, a_{yy}, a_{zz}] = [0, 0.78, 0.84 ] mT$  on the parametric nuclei $D=0$ and $J=0$ at $k_S=10^8$ and $k_T=10^6$ when CISS is only involved in the formation of the radical.}
     \begin{tabular}{||c c c c ||} 
    \hline\hline
    $\chi$ & $ L_1$ & $L_2$ & $L_3$   \\  
     \hline\hline
     $0$ & 7.686209e-01 & 8.887843e-01 & 8.648002e-01 \\
     \hline
      $\frac{\pi}{6}$ & 1.449013e+00 & 1.296792e+00 & 1.117383e+00 \\
     \hline
      $\frac{\pi}{4}$ & 2.135162e+00 & 1.556124e+00  & 1.372103e+00 \\
     \hline
      $\frac{\pi}{3}$ & 2.877980e+00 & 1.746496e+00 & 1.647860e+00 \\
     \hline
      $\frac{\pi}{2}$ & 4.194939e+00 & 1.963228e+00  & 2.136756e+00 \\ 
\hline\hline
\end{tabular}
\end{table}

  \begin{table}
    \centering
    \caption{\label{table2}%
 Number of hyperfine tensors for which $L_1, L_2, L_3$ are greater than unity when each of the components from hyperfine tensor $[a_{xx}, a_{yy}, a_{zz}]$ at $k_S=10^8$ and $k_T=10^6$ when CISS is only involved in the formation of radical pair.}
     \begin{tabular}{||c c c c c||} 
    \hline\hline
     $\chi=0$ & $\chi=\frac{\pi}{6}$  & $\chi=\frac{\pi}{4}$ & $\chi=\frac{\pi}{3}$ & $\chi=\frac{\pi}{2}$    \\  
     \hline\hline
       39 & 1617 &  1241 & 1168 & 1119    \\  
     
\hline\hline
\end{tabular}
\end{table}

It is noteworthy that when CISS exclusively influences the formation of the radical, the observed trend aligns with the experimental findings. A distinctive peak is evident at $500 \mu T$ when CISS is considered. Specifically, when $\chi = \frac{\pi}{2}$, there is a notable increase in the ratio $L_1$ by at least four times. Tab. \ref{table2} provides the count of hyperfine tensors corresponding to the five $\chi$ values that meet the criteria \ref{Experimental_YIeld_Condtions} at  $k_S=10^8$ and $k_T=10^6$, considering CISS solely in the formation of the radical pair.

\subsubsection{CISS at formation and recombination of radical pair }

\begin{figure}
\centering
\includegraphics[width=90mm,keepaspectratio]{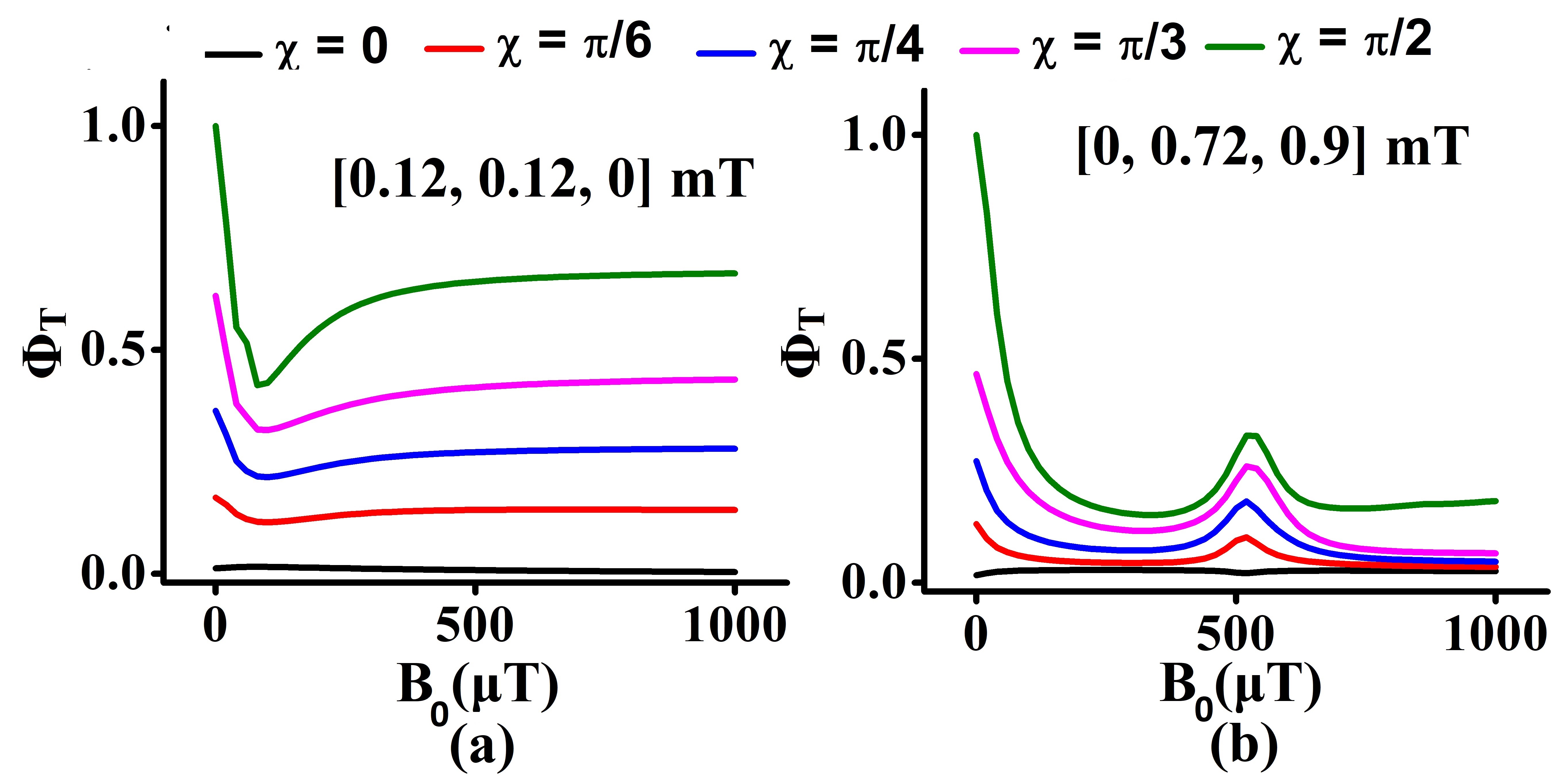}
\caption{Triplet Yield $\phi_T$ with respect to external magnetic field $B_0$  for five values of  $\chi =0, \frac{\pi}{6}, \frac{\pi}{4}, \frac{\pi}{3}, \frac{\pi}{2}$ at  $k_S=10^8$ and $k_T=10^6$ with $D=0$ and $J=0$ at two different hyperfine configuration. (a) $[a_{xx}, a_{yy}, a_{zz}] = [0.12, 0.12, 0 ] mT$ and (b) $[a_{xx}, a_{yy}, a_{zz}] = [0, 0.72, 0.9 ] mT$}
\label{Figure2}
\end{figure}

In Fig.\ref{Figure2}, we plotted triplet yield $\phi_T$ with respect to external magnetic field  for five values of  $\chi =0, \frac{\pi}{6}, \frac{\pi}{4}, \frac{\pi}{3}, \frac{\pi}{2}$ at  $k_S=10^8$ and $k_T=10^6$ with $D=0$ and $J=0$ at two different hyperfine configuration. (a) $[a_{xx}, a_{yy}, a_{zz}] = [0.12, 0.12, 0 ] mT$ and (b) $[a_{xx}, a_{yy}, a_{zz}] = [0, 0.72, 0.9 ] mT$. We list the values of $L_1, L_2, L_3$ in Tab.\ref{table3} and Tab.\ref{table4} for all five values of $\chi$.

\begin{table}
\centering
\caption{\label{table3}%
 $\phi_T$ for radical pair model based on ($O_2^{\do} -Trp^{\dot +}$ with one nuclei. The nuclei have spin one and hyperfine interaction as [$[a_{xx}, a_{yy}, a_{zz}] = [0.12, 0.12, 0 ] mT$]  on the parametric nuclei $D=0$ and $J=0$ at $k_S=10^8$ and $k_T=10^6$ .}
     \begin{tabular}{||c c c c ||} 
    \hline\hline
    $\chi$ & $ L_1$ & $L_2$ & $L_3$   \\  
     \hline\hline
     $0$ & 6.046398e-01 & 5.396299e-01 & 1.120471e+00 \\
     \hline
      $\frac{\pi}{6}$ & 1.129513e+00 & 1.066700e+00 & 1.058885e+00 \\
     \hline
      $\frac{\pi}{4}$ & 1.139740e+00 & 1.076713e+00  & 1.058536e+00 \\
     \hline
      $\frac{\pi}{3}$ & 1.163884e+00 & 1.097928e+00 & 1.060073e+00 \\
     \hline
      $\frac{\pi}{2}$ & 1.189569e+00 & 1.184886e+00 & 1.003952e+00 \\ 
\hline\hline
\end{tabular}
\end{table}

\begin{table}
   \centering
    \caption{\label{table4}%
 $\phi_T$ for radical pair model based on ($O_2^{\do} -Trp^{\dot +}$ with one nuclei. The nuclei have spin one and hyperfine interaction as [$[a_{xx}, a_{yy}, a_{zz}] = [0., 0.72, 0.9 ] mT$]  on the parametric nuclei $D=0$ and $J=0$ at $k_S=10^8$ and $k_T=10^6$ .}
     \begin{tabular}{||c c c c ||} 
    \hline\hline
    $\chi$ & $ L_1$ & $L_2$ & $L_3$   \\  
     \hline\hline
     $0$ & 8.047046e-01 &9.240207e-01 & 8.708729e-01 \\
     \hline
      $\frac{\pi}{6}$ & 2.021043e+00 & 1.196555e+00 & 1.689051e+00 \\
     \hline
      $\frac{\pi}{4}$ & 2.125519e+00 & 1.034806e+00  & 2.054027e+00 \\
     \hline
      $\frac{\pi}{3}$ & 1.688496e+00 & 7.109065e-01 & 2.375131e+00 \\
     \hline
      $\frac{\pi}{2}$ & 1.574199e+00 & 4.780075e-01 & 3.293252e+00 \\ 
\hline\hline
\end{tabular}
\end{table}

 Our parametric analysis found no single combination of hyperfine values for which $L_1, L_2, L_3$ are greater than one for all five values of $\chi$ at $k_S=10^8$ and $k_T=10^6$. The closest result observed was when $[a_{xx}, a_{yy}, a_{zz}] = [0.12, 0.12, 0 ] mT$ when $\chi= \frac{\pi}{6}, \frac{\pi}{4}, \frac{\pi}{3}, \frac{\pi}{2}$ all reported a value of $L_1, L_2, L_3$ greater than 1 however this was not observed for $\chi=0$.
It was observed that the values of $\phi_T$ at $\chi=\frac{\pi}{2}$ are at least ten times larger than that for the case when $\chi=0$. However, the values of the ratio $L_1, L_2, L_3$ remain almost the same. Moreover, the number of hyperfine tensors corresponding to five values of $\chi$ which satisfy criteria \ref{Experimental_YIeld_Condtions} at $k_S=10^8$ and $k_T=10^6$  is given in \ref{table5} when CISS acts both in formation and recombination of radical pair.  \\

  \begin{table}
    \centering
    \caption{\label{table5}%
 Number of hyperfine tensors for which $L_1, L_2, L_3$ are greater than unity when each of the components from hyperfine tensor $[a_{xx}, a_{yy}, a_{zz}]$ at $k_S=10^8$ and $k_T=10^6$ }
     \begin{tabular}{||c c c c c||} 
    \hline\hline
     $\chi=0$ & $\chi=\frac{\pi}{6}$  & $\chi=\frac{\pi}{4}$ & $\chi=\frac{\pi}{3}$ & $\chi=\frac{\pi}{2}$    \\  
     \hline\hline
       39 & 17 &  6 & 2 & 12    \\  
     
\hline\hline
\end{tabular}
\end{table}
Hence, we summarize the difference in the cases when CISS is only present at the formation of the radical (Case A) and when CISS is present both in the formation and recombination of radicals (Case B).
\begin{enumerate}
    \item Case A has a lower value of triplet yield at $45, 200, 500 \mu T$ compared to Case B; hence, case B seems more favorable.
    \item Case A shows an improved increase in the ratio of $L_1, L_2, L_3$ due to CISS compared to Case B at around the same hyperfine tensor; hence, case A seems more favorable.
    \item Case A has more number of hyperfine tensor for $\chi\neq 0$ compared to case B. Since this reaction occurs in a biological system, this property gives the system more degree of freedom and reaction and is not bounded by strict values of hyperfine tensors. Therefore, case A is more promising if we consider this property. 
\end{enumerate}
\subsection{Triplet Yield for varying recombination rates}
\label{RP_CISS_Rate}

In this section, our aim was to investigate the impact of singlet and triplet recombination rates on the yield of the triplet product, $\phi_T$. The goal is to discern a trend that aligns with experimental findings, as articulated in Eq.\ref{Experimental_YIeld_Condtions}, across the maximum number of hyperfine tensors. 
\subsubsection{No CISS}
Recognizing that the planaria reaction transpires in a biological medium, we have exercised the discretion to consider hyperfine tensors. Tabulated in Tab\ref{table6} are the counts of hyperfine configurations, out of the 132,651 possible configurations, that adhere to the observed experimental trend when $\chi=0$.
  \begin{table}
    \centering
    \caption{\label{table6}%
 Number of hyperfine values for which $L_1, L_2, L_3$ are greater than unity when each of the components from hyperfine tensor $[a_{xx}, a_{yy}, a_{zz}]$ is varied from  0 to 3mT such that we cover 132,651 tensor values at $\chi=0$.}
     \begin{tabular}{||c c c c c c||} 
    \hline\hline
    $k_S\downarrow, k_T\rightarrow$ & $ 10^4$ & $10^5$  & $10^6$ & $10^7$ & $10^8$    \\  
     \hline\hline
      $ 10^4$ & 250 & 10 &  90 & 140 & 6    \\  
     \hline
       $ 10^5$ & 315 & 58  & 78 & 135 & 7   \\  
     \hline
       $ 10^6$ & 38 & 103  & 69 & 111 & 7   \\  
     \hline
       $ 10^7$ & 238 & 340  & 223 & 17 & 17    \\  
     \hline
       $ 10^8$ & 2903 & 3227  & 39 & 5 & 6    \\  
\hline\hline
\end{tabular}
\end{table}

Our observation indicates that when $\chi=0$, rate combinations with $ k_S=10^8$ and $k_T=10^5$ exhibit the maximum number of hyperfine configurations satisfying the experimental conditions. However, the absolute value of triplet yield is low due to the diminished triplet recombination rate. Consequently, a higher triplet rate leads to an increased triplet yield value; nevertheless, the rapid recombination rate renders the yield less sensitive to the external magnetic field. This results in a limited number of hyperfine tensor values that meet the specified criteria. Additionally, it was observed that no rate combination resulted in $L_1,L_2,L_3 \ge 2$ (table not shown).
 
\subsubsection{With CISS}

When $\chi=\frac{\pi}{2}$, we have calculated the number of combinations of hyperfine tensors for which $L_1, L_2, L_3$ is greater than unity.  This has been done for case A (Tab.\ref{table7}) and case B (Tab.\ref{table8}) at $\chi=\frac{\pi}{2}$. We generally observe an increase in the number of hyperfine tensors for either case when compared to when $\chi=0$. However, in case A, the number
of hyperfine tensors for which $L_1, L_2, L_3$  is greater than unity is much greater compared to case B for most values of $k_S$ and $k_T$.\\

We have also computed the maximum values of $L_1, L_2, L_3$ when each component of the hyperfine tensor $[a_{xx}, a_{yy}, a_{zz}]$ varies from 0 to 3 mT, covering a total of 132,651 tensor values at $\chi=\frac{\pi}{2}$ for case A (Tab.\ref{tableA3}) and case B (Tab.\ref{tableA1}).
Notably, significant values of $L_1, L_2, L_3$ are observed when $k_T$ is in the range of $10^4$ and $10^5$, with the singlet recombination rate being at least a hundred times ($10^7$ and $10^8$) for both the cases.We also observe that $L_1, L_2, L_3$ values are higher for case A compared to case B. We have also calculated the maximum absolute value of triplet yield (ROS level) at $45\mu T,~ 200 \mu T, ~500 \mu T$ for case A (Tab.\ref{tableA2})  and case B (Tab.\ref{tableA4}) 
It is essential to acknowledge that the rate combination resulting in the maximum value of $L_1, L_2, L_3$ tends to yield lower values of triplet yield compared to yields observed at other rates. This implies that at $\chi=\frac{\pi}{2}$, one can either achieve the maximum value of the ratio $L_1, L_2, L_3$ or attain the absolute yield value by adjusting the recombination rates.

 \begin{table}
    \centering
    \caption{\label{table7}%
 Number of hyperfine values for which $L_1, L_2, L_3$ are greater than unity when each of the components from hyperfine tensor $[a_{xx}, a_{yy}, a_{zz}]$ is varied from  0 to 3mT such that we cover 132,651 tensor values at $\chi=\frac{\pi}{2}$ when CISS only act in formation of radical pair.}
     \begin{tabular}{||c c c c c c||} 
    \hline\hline
    $k_S\downarrow, k_T\rightarrow$ & $ 10^4$ & $10^5$  & $10^6$ & $10^7$ & $10^8$    \\  
     \hline\hline
      $ 10^4$ & 312 & 149 &  59 & 229 & 565   \\  
     \hline
       $ 10^5$ & 15935 & 77  & 51 & 226 & 561   \\  
     \hline
       $ 10^6$ & 14671 & 26423  & 104 & 213 & 556   \\  
     \hline
       $ 10^7$ & 37912 & 22972  & 5268 & 471 & 477    \\  
     \hline
       $ 10^8$ & 10886 & 6692 & 1119 & 426 & 478    \\  
\hline\hline
\end{tabular}
\end{table}

 \begin{table}
    \centering
    \caption{\label{table8}%
 The number of hyperfine values for which $L_1, L_2, L_3$ are greater than unity when each of the components from hyperfine tensor $[a_{xx}, a_{yy}, a_{zz}]$ is varied from  0 to 3mT such that we cover 132,651 tensor values at $\chi=\frac{\pi}{2}$ when CISS acts both in formation and recombination of radical pair.}
     \begin{tabular}{||c c c c c c||} 
    \hline\hline
    $k_S\downarrow, k_T\rightarrow$ & $ 10^4$ & $10^5$  & $10^6$ & $10^7$ & $10^8$    \\  
     \hline\hline
      $ 10^4$ & 8791 & 13086 &  24432 & 18670 & All   \\  
     \hline
       $ 10^5$ & 16775 & 16771  & 26936 & 20022 & 885   \\  
     \hline
       $ 10^6$ & 25311 & 32965  & 24915 & 829 & 0   \\  
     \hline
       $ 10^7$ & 5191 & 4197  & 599 & 0 & 0    \\  
     \hline
       $ 10^8$ & 328 & 191  & 12 & 0 & 0    \\  
\hline\hline
\end{tabular}
\end{table}

\subsection{A higher nuclei study}
\label{RP_CISS_HighNuceli}

This section examines whether the intended experimental outcomes are observed for higher nuclei or if the trend diverges. Given in ~\cite{kinsey2023weak, van2019weak}, indicating that the second nucleus might be tryptophan, we assume the second nucleus to be spin-half, akin to hydrogen. We vary it similarly to a single-nuclei system, but with two nuclei, the number of parameters increases to six. We sampled 51 points from 0 to 3 mT for each component of the hyperfine tensor for a single nuclei system, resulting in 132,651 combinations. This number is $51^6$ tensors for two nuclei systems a very large value. Consequently, we selected values for the hyperfine tensor of the first nucleus that satisfy the conditions outlined in the criteria \ref{Experimental_YIeld_Condtions} as given in the preceding section of one nucleus study and varied the hyperfine tensor for the second nucleus.

\begin{figure}
\centering
\includegraphics[width=90mm,keepaspectratio]{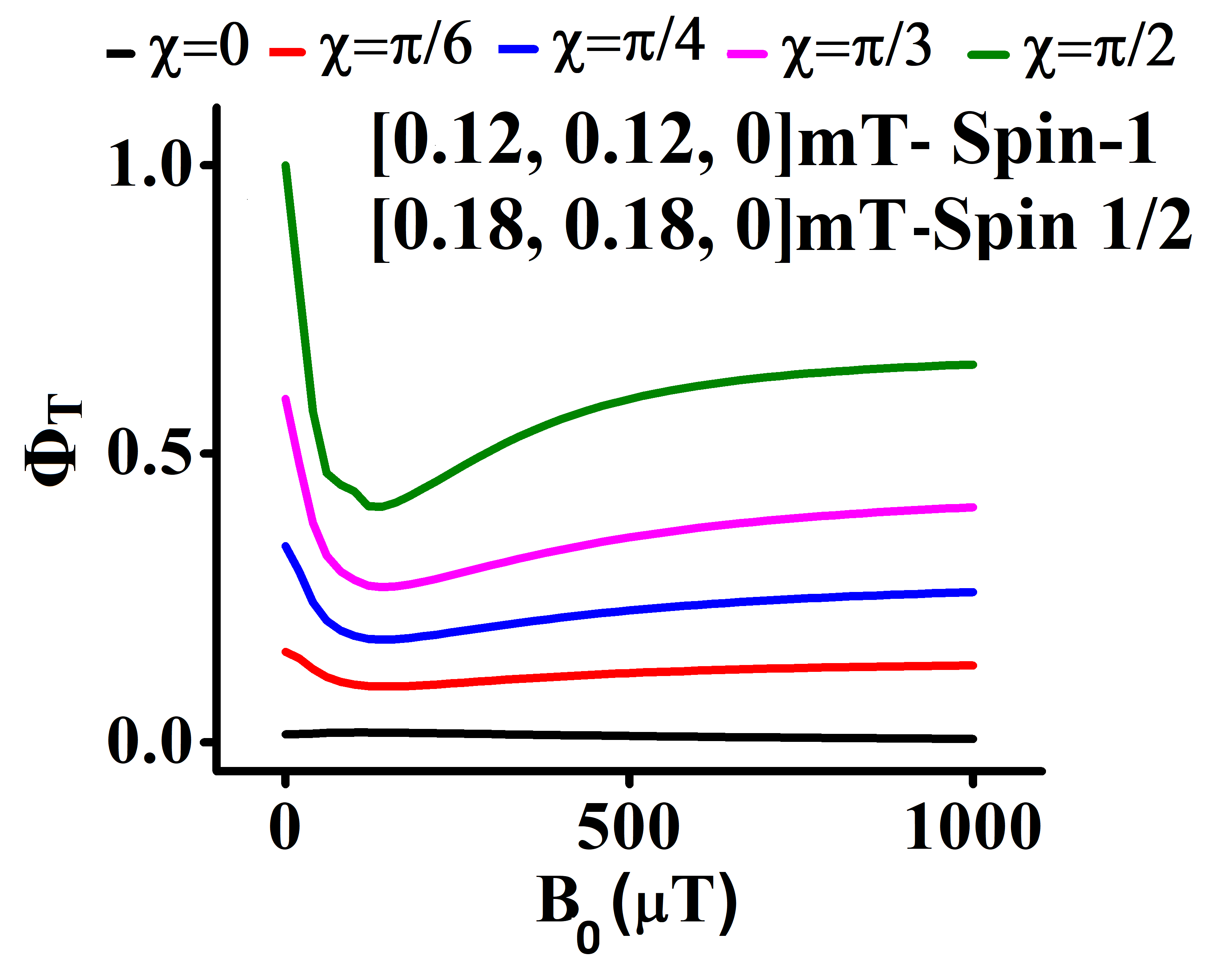}
\caption{Triplet Yield $\phi_T$ with respect to external magnetic field $B_0$  for five values of  $\chi =0, \frac{\pi}{6}, \frac{\pi}{4}, \frac{\pi}{3}, \frac{\pi}{2}$ at  $k_S=10^8$ and $k_T=10^6$ with $D=0$ and $J=0$  at hyperfine configuration $[a_{xx}, a_{yy}, a_{zz}] = [0.12, 0.12, 0 ] mT$ for spin 1 nucleus and $[a_{xx}, a_{yy}, a_{zz}] = [0.18, 0.18, 0 ] mT$ for spin $\frac{1}{2}$  nucleus when CISS is present at formation and recombination. }
\label{Figure7}
\end{figure}

\begin{table}
   \centering
    \caption{\label{table9}%
 $\phi_T$ for radical pair model based on ($O_2^{\do} -Trp^{\dot +}$ with one nuclei. The hyperfine interaction is $[a_{xx}, a_{yy}, a_{zz}] = [0.12, 0.12, 0 ] mT$ for spin 1 nucleus and $[a_{xx}, a_{yy}, a_{zz}] = [0.18, 0.18, 0 ] mT$ for spin $\frac{1}{2}$  nucleus when CISS is present at formation and recombination. at $k_S=10^8$ and $k_T=10^6$ .}
     \begin{tabular}{||c c c c ||} 
    \hline\hline
    $\chi$ & $ L_1$ & $L_2$ & $L_3$   \\  
     \hline\hline
     $0$ & 6.871372e-01 &7.121656e-01 & 9.648560e-01 \\
     \hline
      $\frac{\pi}{6}$ & 1.214984e+00 & 9.440713e-01 & 1.286962e+00 \\
     \hline
      $\frac{\pi}{4}$ & 1.244358e+00 & 9.411223e-01  & 1.322206e+00 \\
     \hline
      $\frac{\pi}{3}$ & 1.275264e+00 & 9.327368e-01 & 1.367228e+00 \\
     \hline
      $\frac{\pi}{2}$ & 1.353438e+00 & 1.037332e+00 & 1.304730e+00 \\ 
\hline\hline
\end{tabular}
\end{table}

We observe in Fig.\ref{Figure7} that for a two nuclei system, we observe the desired trend for $\chi =\frac{\pi}{2}$. The $L_1,L_2,L_3$ are listed in Tab.\ref{table9} for five values of $\chi$. We observe that compared to single nucleus case with hyperfine tensor $[a_{xx}, a_{yy}, a_{zz}] = [0.12, 0.12, 0 ] mT$ Fig.(\ref{Figure2}.a), we observe that $L_1,L_2,L_3$ is greater for $\chi =\frac{\pi}{2}$ for two nuclei case. Hence, we also observe an experimental trend for two nuclei cases. However, due to scaling issues, it might not be possible to give maximum values of $L_1,L_2,L_3$.

Keeping the value of hyperfine tensor fixed $[a_{xx}, a_{yy}, a_{zz}] = [0.12, 0.12, 0 ] mT$, we have done a parametric analysis for the second nucleus and found a number of hyperfine tensors satisfying the criteria for various value of $\chi$ (Tab.\ref{table10}). A similar analysis for hyperfine tensor is$[a_{xx}, a_{yy}, a_{zz}] = [0, 0.72, 0.9 ] mT$ of spin 1 nucleus give us very different results in Tab.\ref{table11}. Here, we obtain no hyperfine tensor for second nuclei, which might satisfy the criteria for $\chi=\frac{\pi}{2}$. Hence, we cannot reach any conclusive result by this brute-force method. 
We also simulated the case when CISS is only present in the formation of the radical pair when the spin one nucleus has hyperfine tensor $[a_{xx}, a_{yy}, a_{zz}] = [0, 0.78, 0.84 ] mT$. We found, in general, that more hyperfine tensors satisfied the criteria (Tab.\ref{table12}). This hints more towards the involvement of CISS only in forming the radical pair. However, a more rigorous approach is required to achieve a more conclusive outcome.

\begin{table}
    \centering
    \caption{\label{table10}%
 Number of hyperfine tensors for which $L_1, L_2, L_3$ are greater than unity when spin 1 nucleus hyperfine tensor is$[a_{xx}, a_{yy}, a_{zz}] = [0.12, 0.12, 0 ] mT$ at $k_S=10^8$ and $k_T=10^6$ We have considered the case when CISS is involved in the formation and recombination of radical pair.}
     \begin{tabular}{||c c c c c||} 
    \hline\hline
     $\chi=0$ & $\chi=\frac{\pi}{6}$  & $\chi=\frac{\pi}{4}$ & $\chi=\frac{\pi}{3}$ & $\chi=\frac{\pi}{2}$    \\  
     \hline\hline
       6814 & 6001 &  2010 & 10 & 50   \\  
     
\hline\hline
\end{tabular}
\end{table}

\begin{table}
    \centering
    \caption{\label{table11}%
 Number of hyperfine tensors for which $L_1, L_2, L_3$ are greater than unity when spin 1 nucleus hyperfine tensor is$[a_{xx}, a_{yy}, a_{zz}] = [0, 0.72, 0.9 ] mT$ at $k_S=10^8$ and $k_T=10^6$ We have considered the case when CISS is involved in the formation and recombination of radical pair.}
     \begin{tabular}{||c c c c c||} 
    \hline\hline
     $\chi=0$ & $\chi=\frac{\pi}{6}$  & $\chi=\frac{\pi}{4}$ & $\chi=\frac{\pi}{3}$ & $\chi=\frac{\pi}{2}$    \\  
     \hline\hline
       11352 & 10 &  1 & 0 & 0    \\  
     
\hline\hline
\end{tabular}
\end{table}

\begin{table}
    \centering
    \caption{\label{table12}%
 Number of hyperfine tensors for which $L_1, L_2, L_3$ are greater than unity when spin 1 nucleus hyperfine tensor is$[a_{xx}, a_{yy}, a_{zz}] = [0, 0.78, 0.84 ] mT$ at $k_S=10^8$ and $k_T=10^6$ We have considered the case when CISS is involved only in formation of radical pair.}
     \begin{tabular}{||c c c c c||} 
    \hline\hline
     $\chi=0$ & $\chi=\frac{\pi}{6}$  & $\chi=\frac{\pi}{4}$ & $\chi=\frac{\pi}{3}$ & $\chi=\frac{\pi}{2}$    \\  
     \hline\hline
       12932 & 21028 &  21005 & 20933 & 20741    \\  
     
\hline\hline
\end{tabular}
\end{table}

\section{Discussion}
The process of blastema formation, crucial for planarian regeneration, seems to be correlated with the level of Reactive Oxygen Species (ROS). Interestingly, the experiments suggest that exposure to Weak Magnetic Fields (WMF) should be continued until the blastema is fully formed.
It's worth noting that the radical pair mechanism, which involves the spin dynamics of electrons and is influenced by magnetic fields, typically has a very short lifetime, on the order of microseconds. This short lifetime raises questions about whether the radical pair mechanism directly participates in the planarian regeneration process or if there's another mechanism at play.
The observed correlation between WMF exposure duration and blastema formation could imply a more complex interplay between magnetic field effects and biological processes. It's possible that the radical pair mechanism, despite its short-lived nature, triggers or influences other biochemical pathways or mechanisms that lead to blastema formation over a longer timescale. Further research is needed to unravel the specifics of these interactions and mechanisms in planarian regeneration.\\
The issue of radical pair formation in the context of planarian regeneration, especially in the absence of an external excitation source, raises intriguing questions. In avian magnetoreception, radical pairs are typically formed through the photoexcitation of neutral acceptor molecules.
In the experiments conducted on planaria regeneration, where Weak Magnetic Fields (WMF) are applied \cite{kinsey2023weak, van2019weak}, there is no apparent external excitation source. The absence of a known external stimulus for radical pair formation prompts the question of what initiates this reaction in the absence of an external excitation source.
One plausible explanation could be related to the cut or segment of the planaria body. The external cut might release some reactive chemical or signaling molecule, which could act as an internal stimulus, leading to the generation of excitation and the subsequent formation of radical pairs in the system. However, the specific identity and nature of this chemical or excitation remain unclear and would require further investigation.
Understanding the underlying biochemical processes triggered by the external cut in planaria regeneration, especially in the context of magnetic field effects, could provide valuable insights into the role of radical pair mechanisms and other potential signaling pathways in this intriguing biological phenomenoa.
\section{Conclusion}

In summary, our analysis suggests that the Radical Pair Mechanism (RPM) provides an explanation for the influence of a weak magnetic field on planaria regeneration. However, our findings indicate that the effect is more robustly explained when considering the Chirality-Induced Spin Selectivity (CISS) in conjunction with the radical pair mechanism. Specifically, when chirality is involved solely in the formation of the radical pair, it yields even more consistent and fitting results based on the experimental findings. This underscores the potential significance of CISS in understanding the magnetic field effects on planaria regeneration with radical pair mechanism.

\begin{acknowledgments}

This work is supported by the Science and Engineering Research Board, Department of Science and Technology (DST), India with grant No. CRG/2021/007060 and DST/INSPIRE/04/2018/000023. 
The authors thank the Department of Electronics and Communication Engineering, IIT Roorkee and the Ministry of Education, Government of India for supporting Y.T.'s graduate research.
\end{acknowledgments}

\appendix
\section{$L_1,L_2,L_3$ and Triplet yield }
\label{Appendix_Hyperfine_Tensors}

In this section, we have listed the  $L_1, L_2, L_3$ at all possible recombination rates of $k_S,k_T$ when CISS is involved in the formation and recombination of radical pair as well as when it is involved only in the formation of the radical. We have also listed the absolute values of the yield at $45\mu T, 200\mu T, 500 \mu T$  for both of the mentioned cases. All the results are for the case when $\chi=\frac{\pi}{2}$.

\begin{table*}
    \centering
    \caption{\label{tableA3}%
 Maximum values of $L_1, L_2, L_3$  when each of the components from hyperfine tensor $[a_{xx}, a_{yy}, a_{zz}]$ is varied from  0 to 3mT such that we cover 132,651 tensor values at $\chi=\frac{\pi}{2}$ when CISS only in formation of radical pair.The maximum value of $L_1, L_2, L_3$ are in general at different hyperfine values }
     \begin{tabular}{||c c c c c c||} 
    \hline\hline
    $k_S\downarrow, k_T\rightarrow$ & $ 10^4$ & $10^5$  & $10^6$ & $10^7$ & $10^8$    \\  
     \hline\hline
      $ 10^4$ & 1.014,1.014,1.001 & 1.0005,1.0005,1.0001 &  1.0002,1.0001,1.0002 & 1.0002,1.0001,1.0002 & 1.0000,1.0000,1.0000   \\  
     \hline
       $ 10^5$ & 1.211,1.189,1.188 & 1.012,1.012,1.011 &  1.002,1.0008,1.002 & 1.002,1.001,1.002 & 1.0001,1.0001,1.000   \\  
     \hline
       $ 10^6$ & 2.58,2.56,1.77 & 1.20,1.19,1.15 &  1.006,1.006,1.006 & 1.017,1.009,1.014 & 1.0007,1.0007,1.0001   \\  
     \hline
       $ 10^7$ & 21.76,20.86,3.91 & 4.06,4.02,2.17   & 1.38,1.34,1.26 & 1.02,1.02,1.006 & 1.004,1.004,1.0007    \\  
     \hline
       $ 10^8$ & 221.61, 70.41, 13.74 & 31.26,10.49,6.87  & 4.19,1.96,2.61  & 1.25,1.13,1.18 &1.0032,1.0032,1.0005    \\  
\hline\hline
\end{tabular}
\end{table*}

 \begin{table*}
    \centering
    \caption{\label{tableA1}%
 Maximum values of $L_1, L_2, L_3$  when each of the components from hyperfine tensor $[a_{xx}, a_{yy}, a_{zz}]$ is varied from  0 to 3mT such that we cover 132,651 tensor values at $\chi=\frac{\pi}{2}$ when CISS only on both formation and recombination of radical pair.The maximum value of $L_1, L_2, L_3$ are in general at different hyperfine values }
     \begin{tabular}{||c c c c c c||} 
    \hline\hline
    $k_S\downarrow, k_T\rightarrow$ & $ 10^4$ & $10^5$  & $10^6$ & $10^7$ & $10^8$    \\  
     \hline\hline
      $ 10^4$ & 1.05,1.04,1.05 & 1.012,1.0097,1.012 &  1.0016,1.0013,1.0012 & 1.0002,1.0001,1.0002 & 1.000,1.000,1.000   \\  
     \hline
       $ 10^5$ & 1.22,1.17,1.20 & 1.05,1.04,1.05  & 1.01,1.01,1.01 & 1.001,1.001,1.001 & 1.0002,1.0001, 1.0002   \\  
     \hline
       $ 10^6$ & 2.65,2.32,1.55 & 1.25, 1.20,1.17  & 1.06,1.04,1.04 & 1.01,1.01,1.01 & $NA$   \\  
     \hline
       $ 10^7$ & 14.95, 10.69, 5.09 & 3.35, 2.52, 2.39   & 1.28, 1.12, 1.22 & $NA$ & $NA$    \\  
     \hline
       $ 10^8$ & 114.17, 18.67, 7.16 & 13.56, 3.13, 5.42  & 1.54, 1.18, 1.46  & $NA$ & $NA$    \\  
\hline\hline
\end{tabular}
\end{table*}

\begin{table*}
\centering
\caption{\label{tableA4}
Maximum values of yield at $45\mu T, 200\mu T, 500 \mu T$  when each of the components from hyperfine tensor $[a_{xx}, a_{yy}, a_{zz}]$ is varied from  0 to 3mT such that we cover 132,651 tensor values at $\chi=\frac{\pi}{2}$ when CISS only in formation of radical pair.The maximum values of yield at $45\mu T, 200\mu T, 500 \mu T$  are in general at different hyperfine values }
     \begin{tabular}{||c c c c c c||} 
    \hline\hline
    $k_S\downarrow, k_T\rightarrow$ & $ 10^4$ & $10^5$  & $10^6$ & $10^7$ & $10^8$    \\  
     \hline\hline
      $ 10^4$  &0.7097,0.7097,0.7131& 0.9569,0.9568,0.9570 &  0.9951,0.9951,0.9952 & 0.9994,0.9994,0.9995 & 0.9999,0.9999,0.9999   \\  
     \hline
       $ 10^5$  &0.312,0.295,0.316 & 0.704,0.704,0.711 &  0.954,0.954,0.954 & 0.994,0.994,0.995 & 0.999,0.999,0.999   
       \\  
     \hline
       $ 10^6$ &  0.21,0.17,0.23 & 0.32,0.32,0.33 &  0.70,0.70,0.71 & 0.947,0.946,0.953 & 0.992,0.992,0.992   \\  
     \hline
       $ 10^7$ & 0.17,0.17,0.17  & 0.25,0.19,0.28  & 0.7,0.34,0.37 & 0.71,0.71,0.71 & 0.93,0.93,0.93    \\  
     \hline
       $ 10^8$ & 0.17,0.17,0.17 & 0.20,0.18,0.23  & 0.38,0.37,0.39 & 0.50,0.49,0.50 & 0.6693,0.6693,0.6693    \\  
\hline\hline
\end{tabular}
\end{table*}

\begin{table*}
\centering
\caption{\label{tableA2}
Maximum values of yield at $45\mu T, 200\mu T, 500 \mu T$  when each of the components from hyperfine tensor $[a_{xx}, a_{yy}, a_{zz}]$ is varied from  0 to 3mT such that we cover 132,651 tensor values at $\chi=\frac{\pi}{2}$ when CISS only on both formation and recombination of radical pair.The maximum value of maximum values of yield at $45\mu T, 200\mu T, 500 \mu T$  are in general at different hyperfine values }
     \begin{tabular}{||c c c c c c||} 
    \hline\hline
    $k_S\downarrow, k_T\rightarrow$ & $ 10^4$ & $10^5$  & $10^6$ & $10^7$ & $10^8$    \\  
     \hline\hline
      $ 10^4$ & 0.89,0.89,0.89 & 0.98,0.98,0.98 &  0.99,0.99,0.99 & 0.99,0.99,0.99 & 1,1,1  \\  
     \hline
       $ 10^5$ & 0.64,0.61,0.64 & 0.89,0.89,0.89  & 0.98,0.98,0.98 & 0.99,0.99,0.99 & 0.99,0.99,0.99   
       \\  
     \hline
       $ 10^6$ & 0.22,0.18,0.32 & 0.64,0.64,0.65 & 0.89,0.89,0.89 & 0.98,0.98,0.98 & $NA$   \\  
     \hline
       $ 10^7$ & 0.37,0.36,0.49 & 0.38,0.38,0.41   & 0.71,0.71,0.73 & $NA$ & $NA$    \\  
     \hline
       $ 10^8$ & 0.24,0.23,0.5 & 0.35,0.31,0.54  & 0.66,0.63,0.67 & $NA$ & $NA$    \\  
\hline\hline
\end{tabular}
\end{table*}

\bibliography{apssamp}

\end{document}